\documentclass[a4paper,fleqn,usenatbib]{mnras}
\usepackage[T1]{fontenc}
\usepackage{ae,aecompl}

\usepackage{graphicx}	
\usepackage{amsmath}	
\usepackage{amssymb}	

\title[Temperature from H~{\sc i} 21 cm absorption studies]{Estimating kinetic temperature from H~{\sc i} 21 cm absorption studies: correction for the turbulence broadening}

\author[Koley \& Roy]{
Atanu Koley$^{1}$\thanks{E-mail: atanuphysics15@gmail.com~(AK)} and Nirupam Roy$^{1}$
\\
$^{1}$Department of Physics, Indian Institute of Science, Bangalore 560012, INDIA
}

\date{Accepted --. Received --; in original form --}

\pubyear{2018}

\begin{document}
\label{firstpage}
\pagerange{\pageref{firstpage}--\pageref{lastpage}}
\maketitle

\begin{abstract}
Neutral hydrogen 21 cm transition is a useful tracer of the neutral 
interstellar medium. However, inferring physical condition from the observed 21 
cm absorption and/or emission spectra is often not straightforward. One 
complication in estimating the temperature of the atomic gas is that the line 
width may have significant contribution from non-thermal broadening. We propose 
a formalism here to separate the thermal and non-thermal broadening using a 
self-consistent model of turbulence broadening of the H~{\sc i} 21 cm 
absorption components. Applying this novel method, we have estimated the spin 
and the kinetic temperature of diffuse Galactic neutral hydrogen, and found 
that a large fraction of gas has temperature in the unstable range. The 
turbulence is found to be subsonic or transonic in nature, and the clouds seem 
to have a bimodal size distribution. Assuming that the turbulence is 
magnetohydrodynamic in nature, the estimated magnetic field strength is of 
$\mu$G order, and is found to be uncorrelated with the H~{\sc i} number 
density.
\end{abstract}




\section{Introduction}

In the thermal steady-state model for neutral hydrogen (H~{\sc i}) in Galactic 
interstellar medium (ISM), two stable phases, the cold neutral medium (CNM; 
kinetic temperature $T_{K}$ $\approx$ 40-200 K) and the warm neutral medium 
(WNM; $T_{K}$ $\geq$ 5000 K), coexist over a narrow range of pressures, 
$P_{min}\leq P \leq P_{max}\approx 3P_{min}$\citep{field65,field69,wolfire95,
wolfire03}. The temperature distribution of the CNM is in good agreement with 
theoretical predictions \citep{clark62,radhakrishnan72,dickey78,heiles03a,roy06}, but, 
due to observational difficulties, little is yet known about the WNM 
\citep{heiles03b,kanekar03,roy13b}. In the CNM, due to higher density ($n 
\approx 10 -100$ cm$^{-3}$), collision is sufficient to thermalize the 
H~{\sc i} 21 cm hyperfine line; thus, the spin temperature ($T_{s}$), which 
basically measures the relative population of the two hyperfine levels, is 
equal to kinetic temperature $T_{K}$. In the WNM, due to low density ($n 
\approx 0.1 -1$ cm$^{-3}$), collision is not so strong to thermalize the 
levels, and hence $T_{s}$ is generally less than $T_{K}$ \citep{liszt01}, 
unless strong Galactic Lyman-$\alpha$ photons thermalize the line 
\citep{field58}. In this simple two phase model, H~{\sc i} at any intermediate 
temperature is expected to be unstable, and drift to either CNM by cooling or 
WNM by heating. But, recently it has been found, both from direct observations 
and realistic simulations, that a significant fraction of the Galactic 
H~{\sc i} has kinetic temperature in the unstable range, $200-5000$ K \citep{heiles03b,kanekar03,roy08,roy13b,murray15,murray18}. Numerical simulations of the ISM 
suggest that turbulence and star formation feedback may play a role in 
redistributing the H~{\sc i} from stable CNM or WNM phase to the thermally 
unstable phase, and the fraction of the unstable gas is strongly correlated with 
the nature of feedback and the strength of the turbulence \citep{audit05,kim14}.

Indeed, measuring the temperature of the diffuse ISM using H~{\sc i}-21 cm line 
has many uncertainties and challenges. Even if the natural width of the line is 
negligible, the broadening has significant non-thermal contribution, and the 
observed linewidth provides only an upper limit to the kinetic temperature 
$T_{K\,max}$. Further, a given line of sight will have multiple components, 
generally though to be isothermal ``clouds''. So, the classical method of 
determining the temperature is to compare the emission and the absorption 
spectra after decomposing them into multiple Gaussian components. Absorption 
spectra are taken towards compact bright continuum sources, whereas emission 
spectra are from nearby lines of sight by assuming that the physical conditions 
are same for both of them \citep{dickey78,payne82,kulkarni88,heiles03b,saha}.Note that 
for the emission spectrum, distribution of gas clouds along the line of sight 
is not known independently, hence it is difficult to decompose the spectrum 
into multiple components including the effect of absorption of background 
components due to optical depth of the foreground ones \citep{heiles03a}. 
Without multi-Gaussian decomposition, we would only infer the column density 
weighted harmonic mean spin temperature of multiple components in a given 
sightline which is biased towards CNM~\citep{roy13b}. Moreover, using this 
method one gets only the spin temperature of individual components, not the 
kinetic temperature. As mentioned earlier, even if $T_{s}$ is coupled to 
$T_{K}$ in the CNM, for the WNM it provides only a lower limit to the kinetic 
temperature. 

\begin{figure}
\includegraphics[width=2.2in,angle=-90]{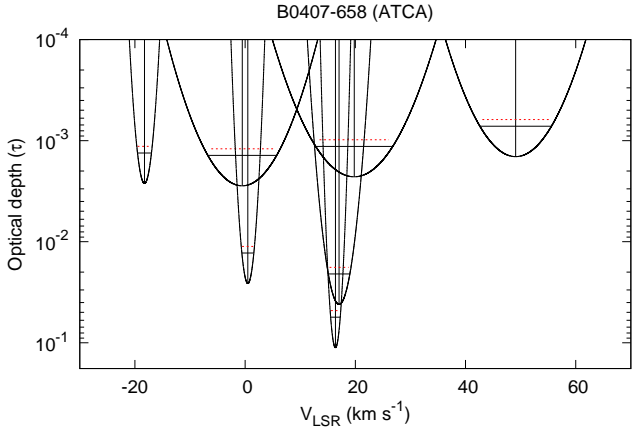}
  \caption{Example best fit multi-Gaussian model spectrum (toward the source B0407-658). For each component, the total and the thermal broadening (full width at half maximum) are shown as the black and the red horizontal lines respectively.}
   \label{fig:fig1}  
\end{figure}

Emission-absorption studies may suffer from further systematic effects if 
either the emission spectra, or both absorption and emission spectra, are from 
single-dish observations. Although we assume that the physical conditions are 
same between the emission and the absorption sightlines, in reality there may 
be small scale variation of H~{\sc i} distribution between the sightlines. The 
analysis will also be affected by sidelobe contamination, spectral baseline 
stability and uncertainty in separating emission and absorption using position 
switching \citep{heiles03b}. Note that, with an interferometer, it is easier 
to reduce or eliminate these systematics, and to measure the absorption spectra 
with high dynamic range to detect the WNM \citep{carilli98,dwarakanath02,roy13b}. 
Of course, the spin temperature measurement will still require observing the 
emission spectra, which is more conveniently done with a single dish telescope.

One way of solving this problem is to use only the more reliable absorption 
spectra to estimate the temperature by separating the thermal and non-thermal 
broadening of the components. To do that, one may use a simple model with some 
scaling relation between the turbulent velocity dispersion ($\sigma_{nth}$) and 
length scale ($l$). For example, incompressible hydrodynamic turbulence follow 
Kolmogorov scaling relation ($\sigma_{nth}$ $\propto$ $l^{1/3}$) \citep{kolmogorov41}. Of course, the ISM is compressible as well as magnetized, so the 
power law index may be different \citep{goldreich95}. However, for the diffuse 
neutral ISM, it is found from observations \citep{roy08,larson79,dutta13}, as well 
as numerical simulation \citep{hennebelle07}, that even if the power law is 
somewhat stepper, it is not very different from a Kolmogorov-like scaling.

In this paper, we have used 21 cm absorption spectra for a sample of Galactic 
lines of sight, and modelled the spectra using a Kolmogorov-like scaling of 
$\sigma_{nth}$ with $l$. Apart from the scaling relation, the method takes into 
consideration the rough thermal pressure equilibrium, and a relation between 
$T_{s}$ and $T_{K}$. In \S2, we have outlined the formalism of how one may 
derive self consistent temperature, density, length scale and column density 
from only the absorption spectra assuming such a scaling law \citep{larson79,
wolfire03} and a model dependent relation between $T_{s}$ and $T_{K}$ \citep{liszt01}. For a {\it consistency check}, the derived column densities are 
compared to the column density estimated from the corresponding emission spectra. 
The analysis and the results are described in \S3, and our main conclusions are 
summarized in \S4.
 
\begin{figure}
\includegraphics[width=2.0in,angle=-90]{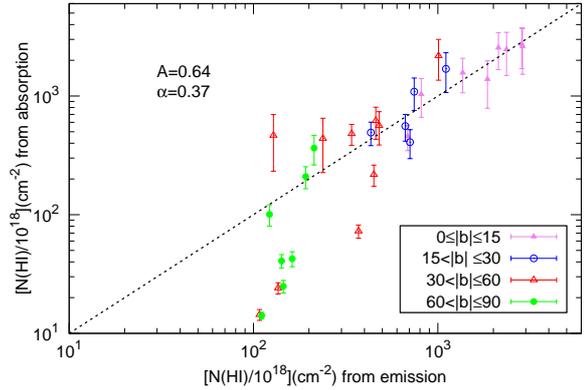}

  \caption{Comparison of estimated column density from the H~{\sc i} emission and absorption spectra. The absorption column density is estimated using the fiducial parameter values of $P = 3700$ Kcm$^{-3}$, $A = 0.64$ and $\alpha = 0.37$. Error bars indicate rms variation for simultaneous variation of P, A and $\alpha$ over the range of 750 - 6000 K/cm$^{3}$, 0.30 - 1.20, and 0.23 - 0.50 respectively.}
   \label{fig:fig2}  
\end{figure}

\begin{figure}
\includegraphics[width=2.0in,angle=-90]{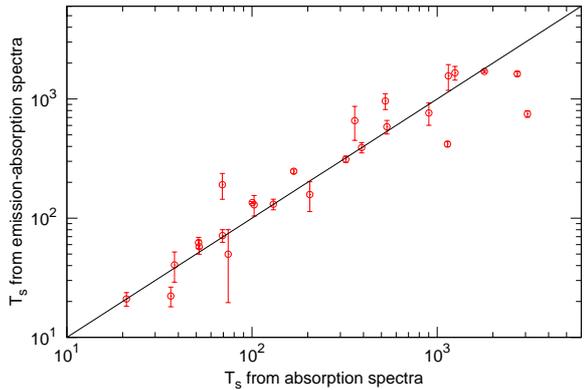}

   \caption{Comparison of estimated spin temperature of individual components from absorption spectra (for $P = 3700$ Kcm$^{-3}$, $A = 0.64$ and $\alpha = 0.37$) and from joint emission-absorption multi-Gaussian decomposition for lines of sight with five or less absorption components.}
    \label{fig:fig3}
\end{figure}

\section{ The Formalism}

Natural line width of H~{\sc i} 21 cm hyperfine line is negligibly small. So the 
line broadening mainly come from the thermal and non-thermal Doppler broadening 
with Gaussian profile. As a result, the observed absorption profile for a 
isothermal component is a Gaussian function with total variance of  
\begin{eqnarray}
\sigma_{total}=(\sigma_{th}^2+\sigma_{nth}^2)^{1/2} \,.
\end{eqnarray}
If an observed absorption profile is fitted with multiple Gaussian components 
with parameters $\tau_{peak}$ (peak optical depth), $\sigma_{total}$ (total 
variance), and $v_{c}$ (center line of sight velocity) for each component, 
these can then be converted to column density under the assumption mentioned 
above. For this, first we separate the thermal and the non-thermal broadening 
in the following way. We start assuming an initial fraction ($0<f<1$) of total 
variance is due to $\sigma_{nth}$, and the rest, from equation (1) is 
$\sigma_{th}$. We then get the kinetic temperature for each component 
\begin{eqnarray}
{T_k}=121~\sigma_{th}^2 \,.
\end{eqnarray}
We assume that the gas is in rough thermal pressure equilibrium, and for a 
constant value of the pressure and typical ISM condition, estimate the 
corresponding $T_{s}$ for each component using results from numerical 
simulations \citep{liszt01}. Note that this relation depends crucially on the 
assumed value of the thermal pressure too. This then allows us to compute the 
column densities of the components 
\begin{eqnarray}
N(H\,I)_{abs}=1.823\times10^{18}\times T_s\, \sqrt\pi \,\tau_{peak}\, b\,.
\end{eqnarray}
From the column density, the kinetic temperature and the assumed value for the 
pressure, we can then estimate the representative length scale of the ``cloud'' 
\begin{eqnarray}
 \displaystyle{l=\frac{N(H\,I)_{abs}{T_k}}{P}}
\end{eqnarray}
which can, then, be used to compute the value of $\sigma_{nth}$ using a scaling 
relation of the form
\begin{eqnarray}
\sigma_{nth}= A~l^{\alpha} \,.
\end{eqnarray}
These calculations are done for an adopted value of $P=3700~Kcm^{-3}$, $A=0.64$ 
and $\alpha=0.37$ \citep{larson79,wolfire03}. The estimated $\sigma_{nth}$ is 
compared with the initial assumed value, and the fraction $f$ is iteratively 
adjusted until a consistent solution is reached. Once the convergence is 
achieved, we get the temperature, density, size and column density for each of 
the components along a line of sight. We then calculate the total column 
density for the line of sight, and, as a consistency check, compare it with the 
total line of sight column density estimated from the emission spectrum. Finally, 
we vary the model parameters $A$ and $\alpha$, and repeat the procedure for 
different values of $P$ to probe how the results are affected by the choice of 
these parameters, and to show that the observed column densities matches more or 
less well for the adopted fiducial values of the parameters.

\begin{figure}
\includegraphics[width=2.2in,angle=-90]{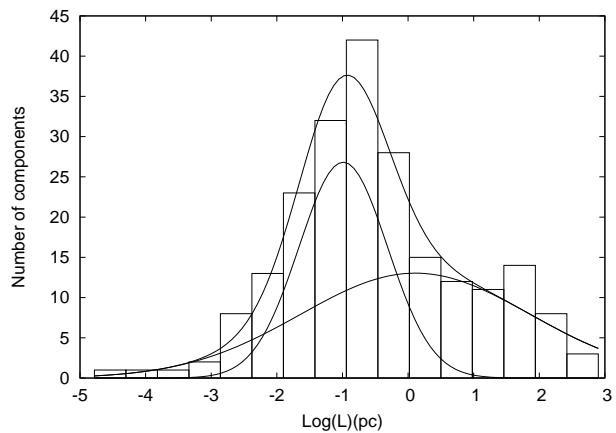}
  \caption{Histogram of the length scale of all 214 components and a two component log-normal fit to the observed distribution for pressure $P=3700~Kcm^{-3}$, $A=0.64$ and $\alpha=0.37$}
   \label{fig:fig4}
\end{figure}

\section{Data Analysis and Results}

All the absorption spectra are taken from the ongoing Galactic H~{\sc i} 21 cm 
absorption line survey \citep{roy13a} carried out with the Giant Metrewave Radio 
Telescope (GMRT), the Westerbork Synthesis Radio Telescope (WSRT), and the 
Australia Telescope Compact Array (ATCA) towards 30 compact radio-loud quasars. 
These 30 lines of sight with 214 components in absorption are used for our 
analysis. The details of the survey and the reduction techniques are discussed 
in details in~\citet{roy13a}. The corresponding emission spectra for overall 
consistency checks are taken from the LAB (Leiden-Argentine Bonn) survey 
\footnote{http://www.astro.uni-bonn.de/en/download/data/lab-survey/} \citep{bajaja05,hartmann97,kalberla05}. Emission column densities were calculated from 
these spectra using the ``isothermal estimation'' - a statistically unbiased 
and more accurate estimate (compared to optically thin estimate) of the 
H~{\sc i} column density - by using the measured brightness temperature from 
LAB survey data and the optical depth from the absorption survey data \citep{chengalur13,roy13b}. The best fit parameters of the Gaussian components (peak 
optical depth, line width and central velocity) are taken from \citet{roy13b}.

\begin{figure}
\includegraphics[width=2.2in,angle=-90]{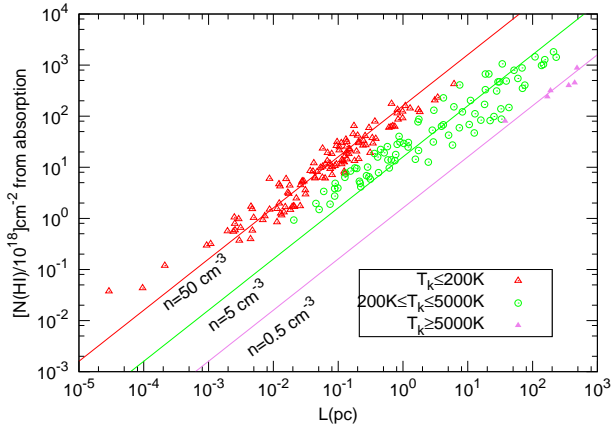}
\caption{Absorption column density as a function of inferred length scale of 214 components. The colours are for different temperature range, and the lines are constant density curves with $n=50$, $5$ and $0.5$ cm$^{-3}$($P=3700~Kcm^{-3}$, $A=0.64$ and $\alpha=0.37$)}
\label{fig:fig5}
\end{figure}

\begin{figure}
\includegraphics[width=2.0in,angle=-90]{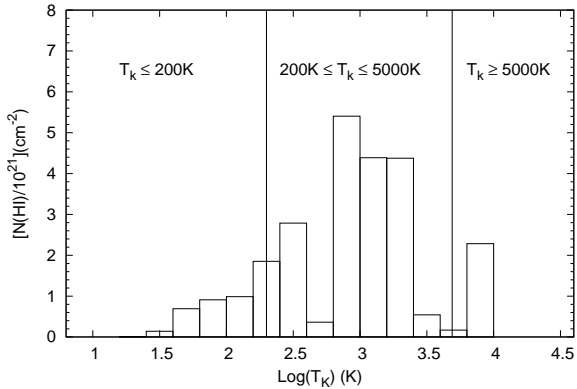}
\caption{H~{\sc i} column density distribution in the cold, warm and unstable phase for $P=3700~Kcm^{-3}$, $A=0.64$ and $\alpha=0.37$.}
\label{fig:fig6}
\end{figure}  
 
We implement the analysis outlined in the previous section for this sample 
through numerical computation. In Figure~\ref{fig:fig1}, we show the result of 
our modeling to separate the thermal and the non-thermal width for an example 
case. The best fit model Gaussian components for the absorption spectra towards 
the source B0407-658 are shown in the figure with the full width half maxima 
(FWHM) and the thermal width as black and red horizontal lines respectively. 
Note that for showing the deep and weak components clearly, the optical depth 
is plotted using logarithmic scale. Clearly, a self-consistent model requires 
non-thermal broadening to explain the observed line width, and this is more 
clearly visible for the wider components.

Figure~\ref{fig:fig2} shows a comparison of the column density using the 
emission and the absorption spectra for the lines of sight of our sample. This 
is for representative values of $P = 3700$ Kcm$^{-3}$, $A=0.64$ and $\alpha = 
0.37$. In general, there is a good match between these two estimations, but for 
six sources, namely B1641+399, B1328+254, B1611+343, B0117-155, B0023-263, 
B0114-211, the emission column density is significantly higher. We note that the 
absorption spectra for these sources at the low H~{\sc i} column density end 
have higher rms noise \citep{roy13b}. Also, four out of these six are lines 
of sight at very high Galactic latitude. It is hence possible that a large 
fraction of the gas is at higher temperature below the detection limit of the 
absorption survey. However, this consistency check indicates that the fiducial 
parameter values adopted allow one to more or less accurately estimate the total 
line of sight column density for this sample solely from the absorption spectra. 
Next, we check the effect of changing various parameters of our model. This is quantified by the change in the estimated column density from the absorption spectra when P, A or $\alpha $ is changed from the adopted values. The estimated column density varies significantly when the parameter values deviate from the adopted fiducial values, and the deviation is larger for a larger variation of the parameter values. This may arguably be an independent
validation of the assumed models, but here we take this as indicative of the fact that the fiducial values adopted for the modeling are more or less the correct values. Hence, for the rest of the analysis, we confined ourselves to these fiducial values only. However, we compute the rms variation of the estimated column density for simultaneous variation of P, A and $\alpha$ uniformly over the range of 750 - 6000 K/cm$^{3}$, 0.30 - 1.20, and 0.23 - 0.50 respectively. This is shown as error bars for each source in Figure ~\ref{fig:fig2}. Note that the mean variation of the estimated column density for the choice of range of the parameters is $\sim 25\%$.

It is assuring that the total column density estimated from only the absorption spectra matches well with the one derived from the emission and absorption spectra ("isothermal" estimate) for each of these lines of sight. But, a more robust method of cross-checking will be a comparison of the estimated spin temperature from this method with the one classically derived from emission-absorption spectra for each individual components. However, that will require joint fitting of the emission and absorption spectra using multiple Gaussian components. Whereas this is already done for the sample of absorption spectra, due to complications mentioned earlier (e.g. uncertain radiative transfer due to self-absorption, relative position of components being unknown, possible stray radiation contamination and relatively unreliable spectral baseline for shallow and wide components), multi-Gaussian decomposition of emission spectrum is not straightforward. Modeling the emission spectra with multiple components for the full sample is beyond the scope of this work, and will be presented for the complete survey (with almost double sample size) in future work. However, for the purpose of comparing the spin temperature of individual components, we have done the fitting for a sub-sample of 10 emission line spectra using data from the LAB survey. These are lines of sight with relatively simpler profile with five or less number of components detected in absorption with low peak optical depth (so that the issue of self absorption is less problematic, and multi-component decomposition is relatively easier and reliable). The fitting of the emission spectra was done with constraints from the absorption spectra in terms of the central velocity of the components, but keeping the amplitude and width as free parameters. Sometime, additional weak and wide components were necessary to achieve a good fit to the data. For those components, all three parameters were kept as free parameters without constraints. For the 25 absorption components along these 10 lines of sight, we derived the spin temperature based on the multi-Gaussian fitting of the emission and absorption spectra. The result is shown in Figure ~\ref{fig:fig3} that compares the spin temperature from this emission-absorption model with that estimated from the absorption spectra only. The match of these two estimation of spin temperature is quite good, and this indicates that the method used for column density estimation is self-consistent and reliable and the adopted fiducial values are more or less the correct values. Hence, for the rest of the analysis, we confined ourselves to these fiducial values only.

\begin{figure}
\includegraphics[width=2.0in,angle=-90]{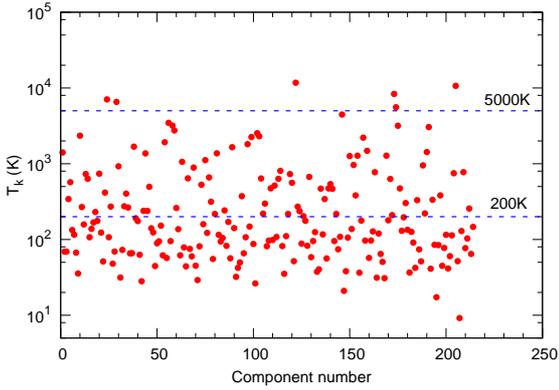}
\caption{Inferred kinetic temperature for all the components for $P=3700~Kcm^{-3}$, $A=0.64$ 
and $\alpha=0.37$.}
\label{fig:fig7}
\end{figure}

\begin{figure}
\includegraphics[width=2.0in,angle=-90]{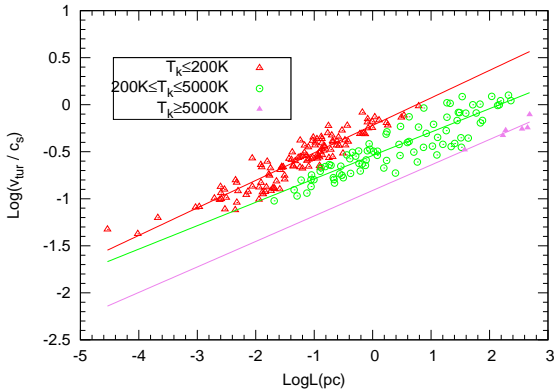}
\caption{The turbulent Mach number as a function of length scale for all components. The colours are for different temperature range, and the lines indicate power law scaling.   ($P=3700~Kcm^{-3}$, $A=0.64$ and $\alpha=0.37$)}
\label{fig:fig8}
\end{figure}

It is interesting to note that the inferred length scales from our analysis, 
shown in Figure~\ref{fig:fig4} have a clear two-component distribution. Fitting 
the observed histogram, we get two log-normal distribution with one peaking at 
about $0.1$ pc and another at $3$ pc with the tail extending as large as few 
hundred pc. The components with smaller length scales have systematically lower 
kinetic temperature and higher density. This is shown in Figure~\ref{fig:fig5} 
where we have overplotted three constant density curves corresponding to $n=50$, 
$5$ and $0.5$ cm$^{-3}$, and the temperature of the components are colour coded 
accordingly. 

In Figure~\ref{fig:fig6}, we show the temperature distribution of the gas based 
on this analysis. We would like to emphasize here, unlike other studies that deal 
with spin temperature (from absorption-emission study) or only the upper limit 
to the kinetic temperature (based on line width), this shows the estimated 
kinetic temperature, albeit certain reasonable assumption, from a self-consistent 
model using only the absorption spectra. The temperature of the individual 
components are shown in Figure~\ref{fig:fig7}. Considering column density 
fraction, we find that about $15\%$ gas is in the cold phase, $\sim 10\%$ gas is 
in the warm phase, and as large as $75\%$ gas has temperature in the intermediate 
range corresponding to the so called unstable phase. Note that the mean kinetic 
temperature of the cold, warm and intermediate phases are $88$, $\sim 8300$ and 
$\sim 940$ K respectively. Interestingly, there are quite a few components where 
the temperature is too low (16 components below $T_{K}\leqslant 40$ K). H~{\sc i} 
with such low temperature has been reported earlier \citep{heiles03b,roy13a}, and 
may be indicative of the absence of small dust grains and polycyclic aromatic 
hydrocarbon, making heating inefficient in some of the compact clouds.
 
\begin{figure}
\includegraphics[width=2.0in,angle=-90]{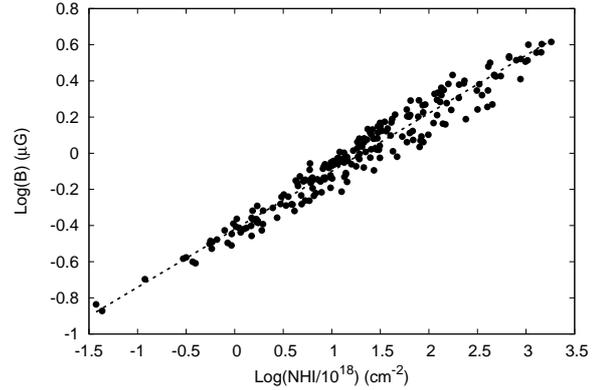}
\caption{Estimated magnetic field and column density for all the components showing a power law scaling with power law index $0.32$. ($P=3700~Kcm^{-3}$, $A=0.64$ and $\alpha=0.37$)}
\label{fig:fig9}
\end{figure} 

\begin{figure}
\includegraphics[width=2.0in,angle=-90]{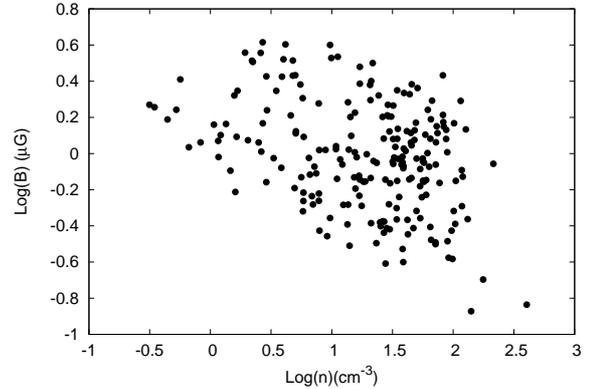}
\caption{Estimated magnetic field and number density for all the components showing no significant correlation.($P=3700~Kcm^{-3}$, $A=0.64$ and $\alpha=0.37$).}
\label{fig:fig10}
\end{figure} 

Next we investigate the strength of turbulence in various phases by computing 
the turbulence Mach number using the estimated temperature, density and the 
turbulent dispersion. Figure~\ref{fig:fig8} shows the estimated Mach number as 
a function of length scale for the different phases. It is found that the 
turbulence is subsonic (and at most transonic) at all scales and all different 
phases. Finally, we estimate the magnetic field for these components under the 
assumption that the turbulence in consideration is magnetohydrodynamic (MHD) in 
nature where Alfven wave is the major energy transfer mode. In that condition, 
the non-thermal velocity dispersion and the magnetic field perturbation 
amplitude are related as $\delta v = \frac{\delta B}{\sqrt{4\pi\mu\ m_{H}\ 
n_{H}}}$, where $\mu$ (mean molecular weight) is 1.4 for H+He and $\delta B 
\approx B$ \citep{aron75,roshi07}. Earlier, \citet{roy08} estimated magnetic field 
using this argument, and the values match with typical diffuse ISM magnetic 
field measured from the Zeeman splitting observations \citep{heiles05}. For 
the current sample also, the magnetic field strength is found to be of the 
order of $\mu$G. Figure~\ref{fig:fig9} shows that the inferred magnetic field 
values and column densities have a power law relation with a power law index of 
$\sim 0.32$, but, as shown in Figure~\ref{fig:fig10}, there is no strong 
correlation of magnetic field and density. This suggests that at densities 
under consideration, field-strength is not increasing significantly due to flux 
freezing.

\section{Conclusions}

Measurement of temperature from H~{\sc i} 21 cm emission and absorption spectra 
is challenging, and has various uncertainties. In this paper, we have outlined a 
method to consistently derive the gas column density and temperature from only 
the absorption spectra, by using a model dependent correction for turbulence 
broadening. This novel formalism is applied to high quality Galactic H~{\sc i} 
absorption spectra for a sub-sample of 30 lines of sight from an ongoing 
absorption line survey. We found that our model, with fiducial scaling relation 
between non-thermal velocity dispersion and length scale, can be used to estimate 
column density, and to infer the column density fraction in different thermal 
phases. This careful analysis establishes, beyond reasonable doubt, the existence 
of a large fraction of gas with the {\it kinetic temperature} in the so called 
unstable range. We also find a bimodal distribution of length scale of the 
absorbing clouds. The non-thermal broadening indicates subsonic or, at most, 
transonic, nature of the turbulence for diffuse neutral ISM. Interestingly, the 
inferred magnetic field strength seems to be increasing monotonically with the 
column density but found to be mostly uncorrelated with the density. We plan to 
apply this analysis for a larger sample from the ongoing absorption survey in 
near future.  

\section*{Acknowledgements}
We thank the anonymous reviewer for useful comments that helped us improve the 
quality of this manuscript significantly. We thank J. N. Chengalur for help, and 
N. Kanekar for useful comments on the manuscript. We are also very grateful to 
Harvey Liszt for providing N.R. his simulation results. This research has made 
use of the NASA$'$S Astrophysics Data System. N.R. acknowledges support from the 
Infosys Foundation through the Infosys Young Investigator grant. A.K. would like to thank DST-INSPIRE (IF160553) for a fellowship. 


\label{lastpage}
\bsp	

\end{document}